\documentstyle[prc,aps,preprint]{revtex}
\begin{document}
\draft
\date{\today}
\title{The role of $\sigma$-meson in $\omega\rightarrow \pi\pi\gamma$ decays \\
and the coupling constant $g_{\omega\sigma\gamma}$}

\author{A. Gokalp~\thanks{agokalp@metu.edu.tr} and
        O. Yilmaz~\thanks{oyilmaz@metu.edu.tr}}
\address{ {\it Physics Department, Middle East Technical University,
06531 Ankara, Turkey}}
\maketitle

\begin{abstract}
We study the $\omega\rightarrow\pi\pi\gamma$ decays by adding to the
amplitude calculated within the framework of chiral perturbation theory
and vector meson dominance the amplitude of  $\sigma$-meson intermediate
state. We estimate the coupling constant $g_{\omega\sigma\gamma}$
utilizing the experimental value of the  $\omega\rightarrow\pi^{0}\pi^{0}\gamma$ 
decay rate.

\end{abstract}

\thispagestyle{empty}
~~~~\\
%\narrowtext
\newpage
\setcounter{page}{1}
%%%
%%%

The radiative decay processes of the type $V\rightarrow PP\gamma$
where P and V belong to the lowest multiplets of vector (V) and
pseudoscalar (P) mesons have been a subject of continuous interest
both theoretically and experimentally. Although the decay rates of
those of these decays that do not involve bremstrahlung radiation
are small and consequently such decays are difficult to detect,
their study nevertheless offers new physics features about the
interesting mechanisms involved in these decays.

Among such radiative $V\rightarrow PP\gamma$ decays, the branching
ratio for the decay $\omega\rightarrow\pi^{0}\pi^{0}\gamma$ has been
measured to be
Br$(\omega\rightarrow\pi^{0}\pi^{0}\gamma)$=$(7.2\pm 2.5)\times 10^{-5}$
\cite{R1}, whereas for the charged mode only on upper limit exists as
Br$(\omega\rightarrow\pi^{+}\pi^{-}\gamma)< 3.6\times 10^{-3}$ \cite{R2}.
Thus by using the well determined $\omega$ full width of
$(8.41\pm0.09)$ MeV \cite{R1}, we obtain for the decay rates of the
radiative decays $\omega\rightarrow\pi^{0}\pi^{0}\gamma$
and $\omega\rightarrow\pi^{+}\pi^{-}\gamma$ the values
$\Gamma^{(exp)}(\omega\rightarrow\pi^{0}\pi^{0}\gamma)=(0.61\pm0.22)$ KeV
and $\Gamma^{(exp)}(\omega\rightarrow\pi^{+}\pi^{-}\gamma)<31$ KeV.

The decay  $\omega\rightarrow\pi\pi\gamma$ as well as other radiative
vector meson decays was first studied by Singer \cite{R3}, who postulated
a mechanism involving the dominance of the intermediate
vector meson contribution (VDM), thus assumed that this decay proceeds
through an intermediate ($\rho\pi$) state as
$\omega\rightarrow (\rho)\pi\rightarrow\pi\pi\gamma$. Singer also
noticed the relation
$\Gamma(\omega\rightarrow\pi^{0}\pi^{0}\gamma)=
\frac{1}{2}\Gamma(\omega\rightarrow\pi^{+}\pi^{-}\gamma)$ for the decay
rates of these reactions calculated using Born-term amplitudes of
VDM contributions where the factor 1/2 is a result of charge
conjugation invariance to order $\alpha$ which imposes pion pairs of even
angular momentum. Renard \cite{R4} later studied radiative decays
$V\rightarrow PP\gamma$ in a gauge  invariant way with current algebra,
hard-pion and Ward-identities techniques. He, moreover, established
the correspondence between these current algebra results and the structure
of the amplitude calculated in the single particle approximation for the
intermediate states. He obtained the value
$\Gamma(\omega\rightarrow\pi^{0}\pi^{0}\gamma)=350$ eV for the contribution
of the VDM amplitude to the decay
$\omega\rightarrow\pi^{0}\pi^{0}\gamma$ which when corrected
by the present day data for the relevant masses and coupling constants
becomes 227 eV \cite{R5}. The contribution  of intermediate vector mesons
to the decay $V\rightarrow PP\gamma$ was also considered
by Bramon et al. \cite{R5} using standard Lagrangians obeying the
SU(3)-symmetry. Their result for the decay rate of the
$\omega\rightarrow\pi^{0}\pi^{0}\gamma$ was
$\Gamma(\omega\rightarrow\pi^{0}\pi^{0}\gamma)=235$ eV.
Fajfer and Oakes \cite{R6} using a low energy effective Lagrangian approach
with gauged Wess-Zumino terms calculated the rate for the decay
$\omega\rightarrow\pi^{0}\pi^{0}\gamma$ as
$\Gamma(\omega\rightarrow\pi^{0}\pi^{0}\gamma)=690$ eV. However,
it should be noted that Bramon et al. \cite{R5} disagree with the numerical
predictions of Fajfer and Oakes \cite{R6} for the decays
$V\rightarrow PP\gamma$ in general, even if the initial expressions for
the Lagrangians are the same. Guetta and Singer \cite{R7} recently
calculated the
$\omega\rightarrow\pi^{0}\pi^{0}\gamma$ decay rate using
the Born amplitude for VDM mechanism as
$\Gamma(\omega\rightarrow\pi^{0}\pi^{0}\gamma)=(344\pm 85)$ eV.
In their calculation they noted that the decay rate of
$\omega\rightarrow\pi^{0}\pi^{0}\gamma$ is proportional to the
coupling constant $g_{\omega\rho\pi}^{2}$ and $g_{\rho\pi\gamma}^{2}$,
and they assumed that the decay $\omega\rightarrow 3\pi$ proceeds with
the same mechanism as $\omega\rightarrow\pi^{0}\pi^{0}\gamma$, that is as
$\omega\rightarrow (\rho)\pi\rightarrow\pi\pi\pi$ \cite{R8}.
They use the experimental input for $\Gamma (\omega\rightarrow 3\pi)$,
$\Gamma (\rho^{0}\rightarrow\pi^{0}\gamma)$, and
$\Gamma (\rho\rightarrow \pi\pi)$ \cite{R2}, and furthermore
they employ a momentum dependent width for $\rho$-meson. If a constant
width is used for $\rho$-meson, then the decay rate is obtained as
$\Gamma (\omega\rightarrow\pi^{0}\pi^{0}\gamma)=306$ eV
using the Born term for the VDM mechanism.

Recently, on the other hand, a new theoretical approach has been developed
for the calculation of $V\rightarrow PP\gamma$ decays within the framework
of chiral effective Lagrangians using chiral perturbation theory \cite{R9}.
Bramon et al. \cite{R9} calculated the decay rates for various decays of
the type $V\rightarrow PP\gamma$ using this approach. They noted that using
chiral perturbation theory Lagrangians there is no tree-level contribution
to the amplitudes for the decay processes $V\rightarrow PP\gamma$ and
that the one-loop contributions are finite and to this order no counterterms
are required. They considered both $\pi\pi$ and $K\bar{K}$ intermediate loops.
In the good isospin limit the $\pi$-loop contributions to the
$\omega\rightarrow\pi^{0}\pi^{0}\gamma$  amplitude vanish and
the contribution of the $K$-loops is very small,
$\Gamma (\omega\rightarrow\pi^{0}\pi^{0}\gamma)_{K}=1.8$ eV.

Since, in addition to the chiral loop contribution, there is always
the term in the decay amplitude given by the intermediate vector meson
dominance (VDM) mechanism \cite{R5,R6}, we obtain for the amplitude
of the decay $\omega\rightarrow\pi^{0}\pi^{0}\gamma$ the theoretical
result
\begin{eqnarray}
A(\omega\rightarrow\pi^{0}\pi^{0}\gamma)=
 A_{\chi}(\omega\rightarrow\pi^{0}\pi^{0}\gamma)
+A_{VDM}(\omega\rightarrow\pi^{0}\pi^{0}\gamma)
\simeq A_{VDM}(\omega\rightarrow\pi^{0}\pi^{0}\gamma)
\end{eqnarray}
where $A_{\chi}$ and $A_{VDM}$ are the chiral and VDM amplitudes, respectively.
However, the central value of the experimental result
$\Gamma(\omega\rightarrow\pi^{0}\pi^{0}\gamma)=0.61$ KeV
for the decay rate of the decay $\omega\rightarrow\pi^{0}\pi^{0}\gamma$ 
is nearly twice the calculated value employing VDM amplitude, hence twice
the theoretical result.

Therefore, the mechanism of the decay
$\omega\rightarrow\pi^{0}\pi^{0}\gamma$ should be reexamined. To this end,
within the theoretical framework of chiral perturbation theory
and vector meson dominance, Guetta and Singer \cite{R7} considered
a neglected feature so far, that is the possibility of $\omega -\rho$ mixing.
However, their calculation showed that the $\omega -\rho$
mixing increases the $\omega\rightarrow\pi^{0}\pi^{0}\gamma$  decay rate
by 5$\%$ only which is even less than 12$\%$ increase provided
by using a momentum dependent width for $\rho$-meson in the calculation
using the VDM amplitude as also noted by these authors.

In this work, we extend our previous studies of the role of $\sigma$-meson
in $\rho^{0}\rightarrow\pi^{+}\pi^{-}\gamma$ \cite{R10} and
$\rho^{0}\rightarrow\pi^{0}\pi^{0}\gamma$ \cite{R11} decays to
$\omega\rightarrow\pi^{0}\pi^{0}\gamma$ and
$\omega\rightarrow\pi^{+}\pi^{-}\gamma$ decays. We follow
a phenomenological approach and attempt to calculate the decay rate
for the decay $\omega\rightarrow\pi^{0}\pi^{0}\gamma$ by considering
$\rho$-pole vector meson dominance amplitude as well as the $\sigma$-pole
amplitude. By employing the experimental value for this decay rate
we calculate the coupling constant $g_{\omega\sigma\gamma}$ as a function
of the experimental $\sigma$-meson parameters
$M_{\sigma}$ and $\Gamma_{\sigma}$ of the $f_{0}(400-1200)$ meson \cite{R2}
which is the candidate for $\sigma$-meson.
We also consider for the $\sigma$ meson parameters the values suggested
by two recent experiments, that is
$M_{\sigma}=555$ MeV $\Gamma_{\sigma}=540$ MeV from CLEO \cite{R20},
and $M_{\sigma}=478$ MeV $\Gamma_{\sigma}=324$ MeV from Fermilab
E791 \cite{R21}.
We then use  the coupling
constant $g_{\omega\sigma\gamma}$ we calculate this way to predict the
decay rate for the $\omega\rightarrow\pi^{+}\pi^{-}\gamma$ decay
assuming that this decay also proceeds through the same mechanism as well,
that is its amplitude is provided by $\rho$-meson and $\sigma$-meson
intermediate state amplitudes. Therefore, the measurement of the
$\omega\rightarrow\pi^{+}\pi^{-}\gamma$ decay rate may provide
us with insight about the mechanism of $V\rightarrow PP\gamma$ decays.

Our calculation is based on the Feynman diagrams shown in Fig. 1 for
$\omega\rightarrow\pi^{0}\pi^{0}\gamma$ decay and on those shown in Fig. 2
for $\omega\rightarrow\pi^{+}\pi^{-}\gamma$ decay. We describe
the $\omega\rho\pi$-vertex by the effective Lagrangian \cite{R12}
\begin{eqnarray}
{\cal L}^{int.}_{\omega\rho\pi}=g_{\omega\rho\pi}
\epsilon^{\mu\nu\alpha\beta}\partial_{\mu}\omega_{\nu}
\partial_{\alpha}\vec{\rho}_{\beta}\cdot\vec{\pi}
\end{eqnarray}
which also conventionally defines the coupling constant $g_{\omega\rho\pi}$.
Since there is no phase space to measure an $\omega\rightarrow\rho\pi$
transition, this vertex should be extracted from theoretical models.
Vector Meson Dominance and current-field identities \cite{R13} gives
$g_{\omega\rho\pi}\simeq 12$ GeV$^{-1}$ while approximate SU(3)
symmetry suggests  $g_{\omega\rho\pi}\simeq 16$ GeV$^{-1}$ \cite{R14}.
On the other hand, QCD sum rule calculations obtain the value
$g_{\omega\rho\pi}\simeq (15-17)$ GeV$^{-1}$ \cite{R12}, and the
light cone QCD sum rules method extracts the value
$g_{\omega\rho\pi}= 15$ GeV$^{-1}$ \cite{R15}. Recently QCD sum
rules method for the polarization operator in an external field yields
the value $g_{\omega\rho\pi}\simeq 16$ GeV$^{-1}$ \cite{R16}. In this work we use
this coupling constant as $g_{\omega\rho\pi}=15$ GeV$^{-1}$.
For the $\sigma\pi\pi$-vertex we use the effective Lagrangian \cite{R17}
\begin{eqnarray}
{\cal L}^{int.}_{\sigma\pi\pi}=\frac{1}{2}g_{\sigma\pi\pi}M_{\sigma}
\vec{\pi}\cdot\vec{\pi}\sigma~~.
\end{eqnarray}
The decay width of the $\sigma$-meson that follows from this effective
Lagrangian is given as
\begin{eqnarray}
\Gamma_{\sigma}\equiv\Gamma(\sigma\rightarrow\pi\pi)=
\frac{g^{2}_{\sigma\pi\pi}}{4\pi}\frac{3M_{\sigma}}{8}
\left [ 1-(\frac{2M_{\pi}}{M_{\sigma}})^{2}\right ] ^{1/2}~~.
\end{eqnarray}
The $\rho\pi\gamma$-vertex is described by the effective Lagrangian \cite{R13}
\begin{eqnarray}
{\cal L}^{int.}_{\rho\pi\gamma}=\frac{e}{M_{\rho}}g_{\rho\pi\gamma}
\epsilon^{\mu\nu\alpha\beta}\partial_{\mu}\vec{\rho}_{\nu}\cdot\vec{\pi}
\partial_{\alpha}A_{\beta}~~.
\end{eqnarray}
The coupling constant $g_{\rho\pi\gamma}$ can then be obtained from the
experimental partial width of the radiative decay $\rho\rightarrow\pi\gamma$
\cite{R2}. However, at present there appears to be a discrepancy between the
experimental widths of the $\rho^{0}\rightarrow\pi^{0}\gamma$ and
$\rho^{+}\rightarrow\pi^{+}\gamma$ decays \cite{R2}. We use the experimental
rate for the decay $\rho^{0}\rightarrow\pi^{0}\gamma$  to extract
the coupling constant $g_{\rho\pi\gamma}$ as $g_{\rho\pi\gamma}^{2}=0.485$
since in our calculation we use the experimental value for the decay rate of
$\omega\rightarrow\pi^{0}\pi^{0}\gamma$ to estimate the coupling constant
$g_{\omega\sigma\gamma}$. Finally, we describe the
$\omega\sigma\gamma$-vertex by the effective Lagrangian \cite{R18}
\begin{eqnarray}
{\cal L}^{int.}_{\omega\sigma\gamma}=\frac{e}{M_{\omega}}g_{\omega\sigma\gamma}
   [\partial^{\alpha}\omega^{\beta}\partial_{\alpha}A_{\beta}
   -\partial^{\alpha}\omega^{\beta}\partial_{\beta}A_{\alpha}]\sigma~~,
\end{eqnarray}
which also defines the coupling constant $g_{\omega\sigma\gamma}$.

In our calculation of the invariant amplitudes for the decays
$\omega\rightarrow\pi^{0}\pi^{0}\gamma$  and
$\omega\rightarrow\pi^{+}\pi^{-}\gamma$, in the $\sigma$-meson propagator
we make the replacement
$M_{\sigma}\rightarrow M_{\sigma}-\frac{1}{2}i\Gamma_{\sigma}$ , where
$\Gamma_{\sigma}$ is given by Eq. 4. Since the experimental candidate
for $\sigma$-meson $f_{0}(400-1200)$ has a width of (600-1000) MeV \cite{R2},
we estimate the coupling costant $g_{\omega\sigma\gamma}$ from the
experimental decay rate of the $\omega\rightarrow\pi^{0}\pi^{0}\gamma$
decay for a set of values of $\sigma$-meson parameters $M_{\sigma}$ and
$\Gamma_{\sigma}$.
We furthermore consider the results of two recent experiments for $\sigma$
meson parameters \cite{R20}-\cite{R21}.
On the other hand, for $\rho$-meson propagator
we also make the replacement
$M_{\rho}\rightarrow M_{\rho}-\frac{1}{2}i\Gamma_{\rho}$  but
use the constant experimental width of $\rho$-meson.

In terms of the invariant amplitude ${\cal M}$(E$_{\gamma}$, E$_{1}$),
the differential decay probability  of
$\omega\rightarrow\pi\pi\gamma$ decay for an unpolarized
$\omega$-meson at rest is then given as
\begin{eqnarray}
\frac{d\Gamma}{dE_{\gamma}dE_{1}}=\frac{1}{(2\pi)^{3}}~\frac{1}{8M_{\omega}}~
\mid {\cal M}\mid^{2} ,
\end{eqnarray}
where E$_{\gamma}$ and E$_{1}$ are the photon and pion energies respectively.
We perform an average over the spin states of
$\omega$-meson and a sum over the polarization states of the photon.
The decay width $\Gamma(\omega\rightarrow\pi\pi\gamma)$ is then
obtained by integration
\begin{eqnarray}
\Gamma=(\frac{1}{2})\int_{E_{\gamma,min.}}^{E_{\gamma,max.}}dE_{\gamma}
       \int_{E_{1,min.}}^{E_{1,max.}}dE_{1}\frac{d\Gamma}{dE_{\gamma}dE_{1}}
\end{eqnarray}
where now the factor ($\frac{1}{2}$) is included in the calculation of
the width of the $\omega\rightarrow\pi^{0}\pi^{0}\gamma$ decay because
of the $\pi^{0}\pi^{0}$ pair in the final state.
The minimum photon energy is E$_{\gamma, min.}=0$ and the maximum photon
energy is given as
$E_{\gamma,max.}=(M_{\omega}^{2}-4M_{\pi}^{2})/2M_{\omega}$=341 MeV.
The maximum and minimum values for pion energy E$_{1}$ are given by
\begin{eqnarray}
\frac{1}{2(2E_{\gamma}M_{\omega}-M_{\omega}^{2})}
[ -2E_{\gamma}^{2}M_{\omega}+3E_{\gamma}M_{\omega}^{2}-M_{\omega}^{3}
 ~~~~~~~~~~~~~~~~~~~~~~~~~~~~ \nonumber \\
\pm  E_{\gamma}\sqrt{(-2E_{\gamma}M_{\omega}+M_{\omega}^{2})
       (-2E_{\gamma}M_{\omega}+M_{\omega}^{2}-4M_{\pi}^{2})}~] ~.
\nonumber
\end{eqnarray}

We first consider the $\omega\rightarrow\pi^{0}\pi^{0}\gamma$ decay
and using the experimental value for its decay rate we estimate the
coupling constant $g_{\omega\sigma\gamma}$. Since the theoretical decay
rate we calculate using Feynman diagrams in Fig. 1 results
in a quadric equation for the coupling constant $g_{\omega\sigma\gamma}$,
for a given set of $\sigma$-meson parameters $M_{\sigma}$ and $\Gamma_{\sigma}$
we obtain two values for this coupling constant, one being positive
and one being negative. We present the results of our calculation in
the first two columns of Table 1.
We then consider the $\omega\rightarrow\pi^{+}\pi^{-}\gamma$ decay and
using the experimental upper limit for its decay rate and the
theoretical value we calculate from Feynman diagrams in Fig. 2
this time we obtain upper and lower limits, in other words an interval,
for the coupling constant $g_{\omega\sigma\gamma}$. The results
are presented in the last two columns of Table 1 again for
different sets of parameters $M_{\sigma}$ and $\Gamma_{\sigma}$.
These results determine an interval for $g_{\omega\sigma\gamma}$,
for example for $M_{\sigma}=500$ MeV and $\Gamma_{\sigma}=600$ MeV
the interval is $-1.73<g_{\omega\sigma\gamma}<1.58$.
Examination of the results in the first two and the last two columns
of Table 1 show that these results are consistent with each other.
We then, by using the values of the coupling constant $g_{\omega\sigma\gamma}$
we estimate from the $\omega\rightarrow\pi^{0}\pi^{0}\gamma$ decay,
calculate the decay rate of $\omega\rightarrow\pi^{+}\pi^{+}\gamma$ decay
using a given set of $\sigma$-meson parameters
$M_{\sigma}$ and $\Gamma_{\sigma}$.
We like to note, however, that our results for the
$\omega\rightarrow\pi^{+}\pi^{-}\gamma$ decay rate does not change
for different set of $M_{\sigma}$ and $\Gamma_{\sigma}$ as expected, so that we
only give the result for the set $M_{\sigma}=478$ MeV and
$\Gamma_{\sigma}=374$ MeV obtained in the recent Fermilab experiment E791
\cite{R21}. If we choose the positive value for
$g_{\omega\sigma\gamma}=0.13$ the resulting branching ratio for the decay
$\omega\rightarrow\pi^{+}\pi^{-}\gamma$ is
$Br(\omega\rightarrow\pi^{+}\pi^{-}\gamma)=14.6\times 10^{-5}$ and
for $g_{\omega\sigma\gamma}=-0.27$ it is also
$Br(\omega\rightarrow\pi^{+}\pi^{-}\gamma)=14.6\times 10^{-5}$ in accordance with Singer's
theorem \cite{R3}.

The photon spectra for the decay rate of the decay
$\omega\rightarrow\pi^{0}\pi^{0}\gamma$ are plotted in Fig. 3 and
in Fig. 4.
In these figures we use the $\sigma$-meson parameters
$M_{\sigma}=478$ MeV, $\Gamma_{\sigma}=324$ MeV resulting in
the coupling constant $g_{\sigma\pi\pi}=5.29$. In Fig. 3 we use
the positive value of the coupling constant
$g_{\omega\sigma\gamma}=0.13$ and in Fig. 4 the negative value
$g_{\omega\sigma\gamma}=-0.27$.
The general shape of the spectrum as well as the relative contributions of
different terms for positive and negative values of $g_{\omega\sigma\gamma}$ are
quite different. These figures clearly show the importance of
the $\sigma$-amplitude term and of the interference term between $\sigma$-amplitude
and VDM-amplitude. As a matter of fact for these values of
$M_{\sigma}$ and $\Gamma_{\sigma}$, and for $g_{\omega\sigma\gamma}=0.13$
we obtain for the decay rate of $\omega\rightarrow\pi^{0}\pi^{0}\gamma$
calculated using
$\sigma$- and VDM-amplitudes the results
$\Gamma_{VDM}(\omega\rightarrow\pi^{0}\pi^{0}\gamma)=291$ eV and
$\Gamma_{\sigma}(\omega\rightarrow\pi^{0}\pi^{0}\gamma)=156$ eV,
and the interference term between $\sigma$- and VDM-amplitudes contributes
$\Gamma_{inter}(\omega\rightarrow\pi^{0}\pi^{0}\gamma)=167$ eV to the
decay rate. On the other hand for the negative value of the coupling constant
$g_{\omega\sigma\gamma}=-0.27$
we obtain for the decay rate of $\omega\rightarrow\pi^{0}\pi^{0}\gamma$
calculated using
$\sigma$- and VDM-amplitudes the results
$\Gamma_{VDM}(\omega\rightarrow\pi^{0}\pi^{0}\gamma)=291$ eV and
$\Gamma_{\sigma}(\omega\rightarrow\pi^{0}\pi^{0}\gamma)=674$ eV,
and the interference term between $\sigma$- and VDM-amplitudes now makes a negative
contribution $\Gamma_{inter}(\omega\rightarrow\pi^{0}\pi^{0}\gamma)=-347$ eV to the
decay rate.

We like to stress that by considering the contribution of the $\sigma$-meson
intermediate state to the amplitudes for the $\omega\rightarrow\pi\pi\gamma$
decays we may resolve the discrepancy between the experimental value
and the theoretical result calculated within the framework of chiral
perturbation theory and Vector Meson Dominance for the
$\omega\rightarrow\pi^{0}\pi^{0}\gamma$ decay rate. Moreover,
since we also make prediction for the photon spectra of
$\omega\rightarrow\pi^{+}\pi^{-}\gamma$ decay
its measurement can provide a test for the mechanism of
$\omega\rightarrow\pi\pi\gamma$ decays, and the sign of the coupling constant
$g_{\omega\sigma\gamma}$.

Furthermore, the coupling constant $g_{\omega\sigma\gamma}$ is an important
physical input for studies of $\omega$-meson photoproduction on nucleons
\cite{R19}. Although at sufficiently high energies and low momentum transfers
electromagnetic production of vector mesons on nucleon targets has been
explained by Pomeron exchange models, at low energies near threshold scalar
and pseudoscalar meson exchange mechanisms becomes important \cite{R18},
so that the coupling constant $g_{\omega\sigma\gamma}$ is required for the
analysis of photoproduction reactions of $\omega$-meson on nucleons near
threshold within the framework of meson-exchange mechanism.

\begin{center}
{\bf ACKNOWLEDGMENTS}
\end{center}

We thank  M. P. Rekalo for suggesting this problem to us and for
his interest during the course of our work. We also like to thank
P. Singer for drawing our attention to an error in the preprint of our paper.

%\pagebreak

\begin{table}
\caption{The coupling constant $g_{\omega\sigma\gamma}$ for different set of
$\sigma$-meson parameters. First two columns show $g_{\omega\sigma\gamma}$ estimated from
the experimental rate of the $\Gamma(\omega\rightarrow \pi^{0}\pi^{0}\gamma)$ decay,
and the last two columns show the upper and lower limits for $g_{\omega\sigma\gamma}$
estimated from the experimental upper limit of the
$\Gamma(\omega\rightarrow \pi^{+}\pi^{-}\gamma)$ decay rate.}
\begin{tabular}{|p{0.7in}|p{0.4in}||c|c|p{0.5in}||p{0.5in}|p{0.4in}|}
\multicolumn{2}{|c||}{$g_{\omega\sigma\gamma}(\omega\rightarrow\pi^{0}\pi^{0}\gamma)$}
&$M_{\sigma}$ (MeV)
&$\Gamma_{\sigma}$ (MeV)
&$g_{\sigma\pi\pi}$
&\multicolumn{2}{c|}{$g_{\omega\sigma\gamma}(\omega\rightarrow\pi^{+}\pi^{-}\gamma)$}
\\
\hline
~~~  0.18 & -0.33 & 500 & 600 & 6.97 & 1.58 & -1.73 \\ \hline
~~~  0.21 & -0.36 & 500 & 800 & 8.04 & 1.80 & -1.95 \\ \hline
~~~  0.24 & -0.43 & 600 & 800 & 7.11 & 2.10 & -2.30 \\ \hline
~~~  0.27 & -0.51 & 700 & 800 & 6.46 & 2.45 & -2.69 \\ \hline
~~~  0.30 & -0.59 & 800 & 600 & 5.18 & 2.73 & -3.02 \\ \hline
~~~  0.32 & -0.61 & 800 & 900 & 6.34 & 2.91 & -3.19 \\ \hline
~~~  0.37 & -0.71 & 900 & 900 & 5.94 & 3.31 & -3.65 \\ \hline
~~~  0.18 & -0.36 & 555 & 540 & 6.15 & 1.68 & -1.85 \\ \hline
~~~  0.13 & -0.27 & 478 & 324 & 5.29 & 1.20 & -1.34 \\
\end{tabular}
\end{table}

\newpage

{\bf Figure Captions:}

\begin{description}

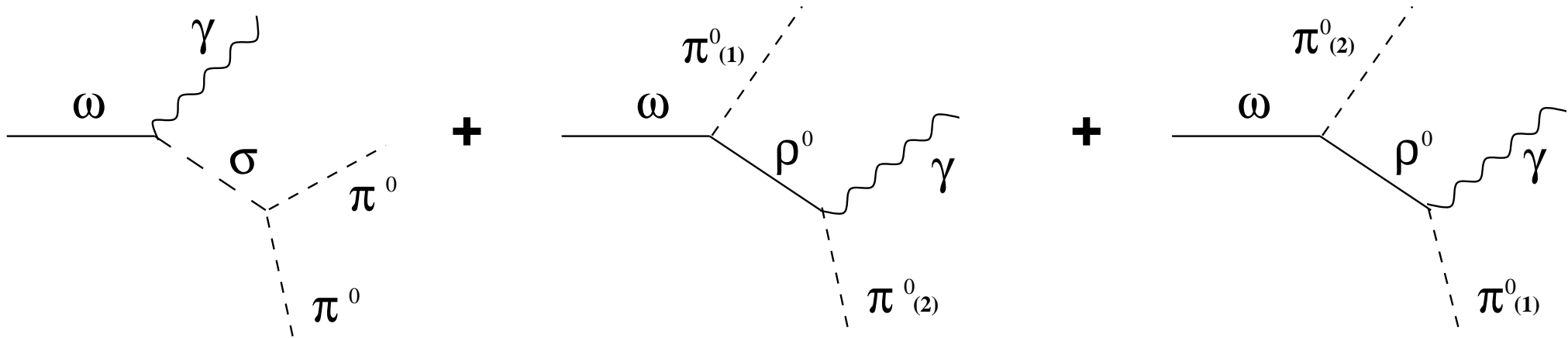
\item[{\bf Figure 1}:] Feynman Diagrams for the decay
$\omega\rightarrow \pi^{0}\pi^{0}\gamma$

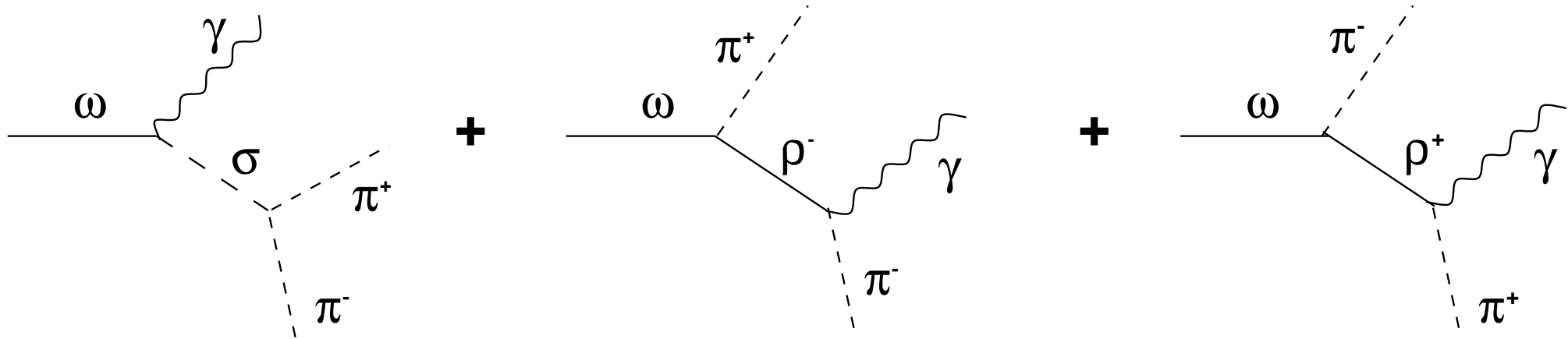
\item[{\bf Figure 2}:] Feynman Diagrams for the decay
$\omega\rightarrow \pi^{+}\pi^{-}\gamma$

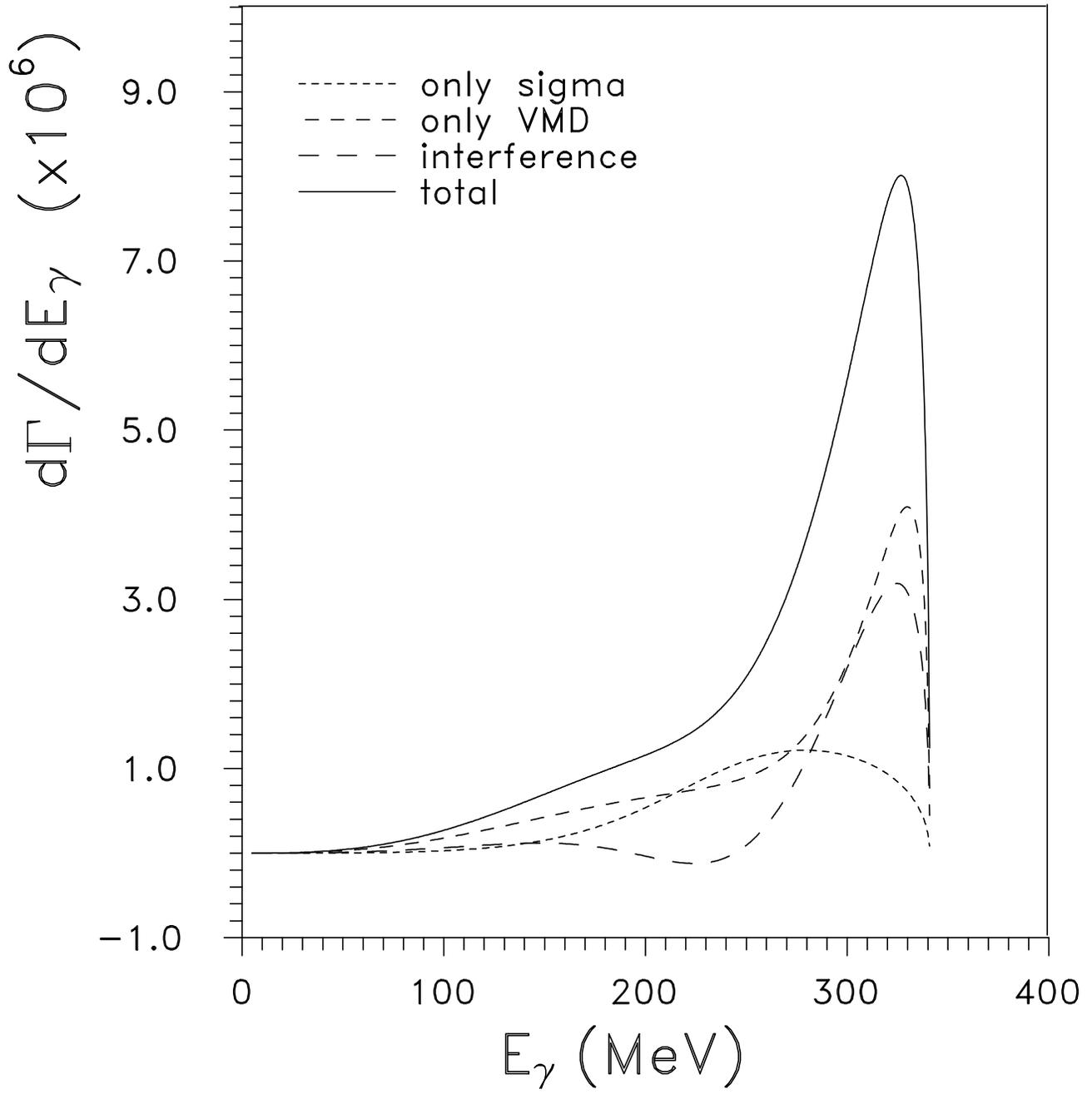
\item[{\bf Figure 3}:] The photon spectra for the decay width of
$\omega\rightarrow\pi^{0}\pi^{0}\gamma$ for $g_{\omega\sigma\gamma}>0$.
The contributions of different terms are indicated.

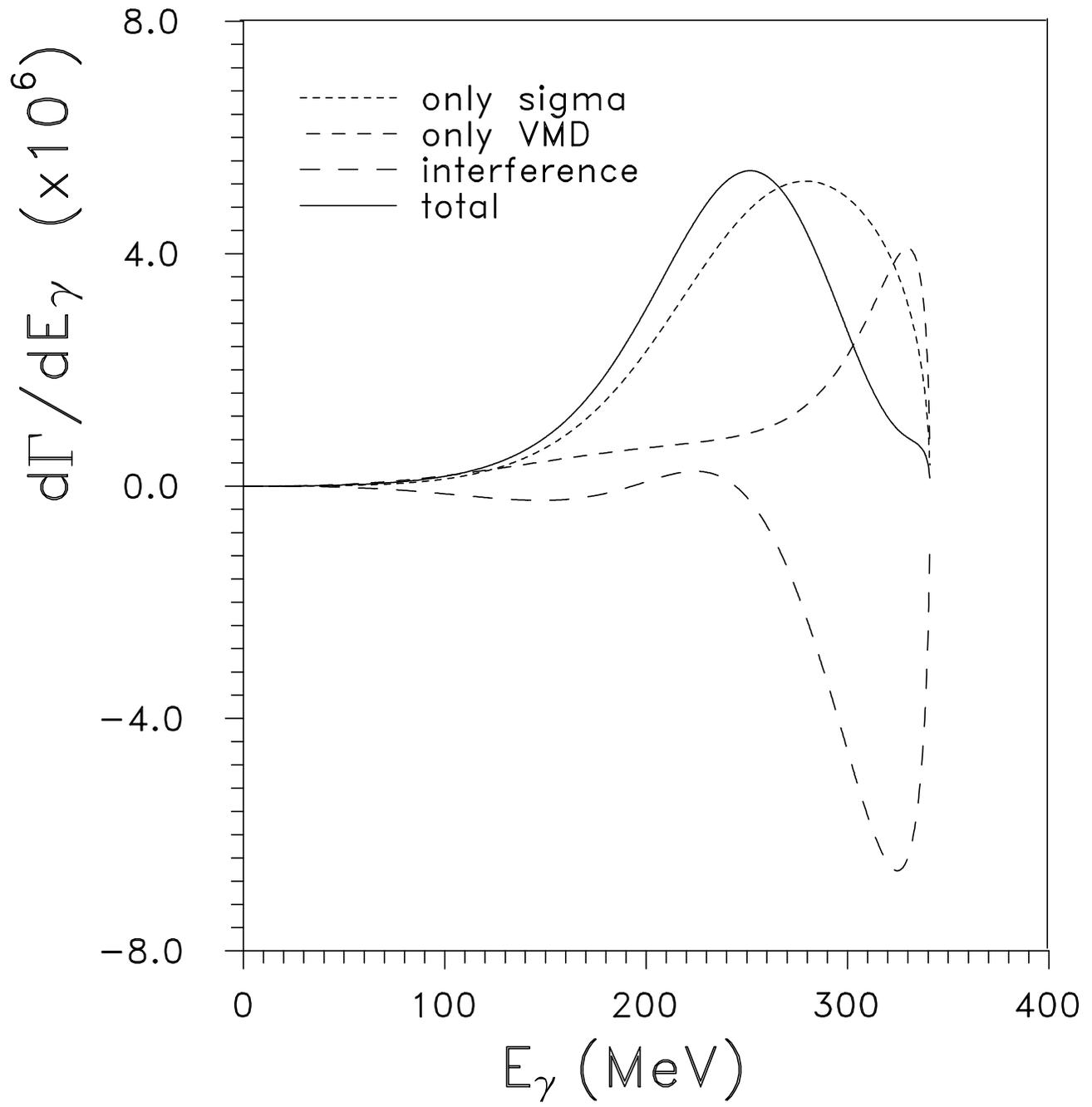
\item[{\bf Figure 4}:] The photon spectra for the decay width of
$\omega\rightarrow\pi^{0}\pi^{0}\gamma$ for $g_{\omega\sigma\gamma}<0$.
The contributions of different terms are indicated.

\end{description}

\end{document}